\documentclass[letterpaper]{article} 

\usepackage{chapterbib}
\usepackage{graphicx}
\usepackage{dcolumn}
\usepackage{bm}
\usepackage{amsmath, amssymb} 
\usepackage{hyperref}   
\usepackage[american]{babel}
\usepackage{siunitx}    
\usepackage{pdflscape}
\usepackage[T1]{fontenc}
\usepackage{float}
\usepackage{authblk}
\usepackage{multirow}
\usepackage{titlesec}
\usepackage[letterpaper, left=1.2in, right=1.2in, top=1in, bottom=1in]{geometry}
\usepackage[articletitle=true]{achemso}

\begin{document}

\title{High-Throughput-Screening Workflow for Predicting Volume Changes by Ion Intercalation in Battery Materials}

\author[1,2]{Aljoscha Felix Baumann *}
\affil[1]{Fraunhofer IWM,  W\"ohlerstra{\ss}e 11, 79108 Freiburg, Germany}
\affil[2]{Freiburg Materials Research Center (FMF), Albert-Ludwigs-University of Freiburg, Stefan-Meier-Stra{\ss}e 21, 79104 Freiburg, Germany}
\author[1]{Daniel Mutter}
\author[1,2]{Daniel F. Urban}
\author[1,2]{Christian Els\"asser}

\date{aljoscha.baumann@iwm.fraunhofer.de}

\maketitle

\begin{abstract}

Mechanical stresses and strains developing locally within the microstructure of active ion-battery-electrode materials during charge-discharge cycles can compromise their long-term stability. In this context, crystalline compounds exhibiting low volume changes are of particular interest. Atomistic simulations can be employed to quantify the volume change of the crystal structure upon intercalation and deintercalation of ions and to elucidate the local mechanisms underlying the global structural response. While density functional theory (DFT) offers a robust and accurate framework for such calculations, its computational cost limits its applicability for large-scale screening of diverse intercalation structures and sites. In this work, we present a workflow designed to prioritize candidate materials for subsequent detailed characterization. The workflow calculates the volume change upon intercalation using atomic-level features and a machine-learning model for bond-length prediction. The bond-length predictions are based on the assumption that bonds between the same ionic species in similar local coordination environments exhibit comparable lengths across different crystallographic structures. The model was trained on a DFT-generated dataset, which inherently defines the chemical space in which reliable predictions can be expected. We demonstrate the workflow's utility by screening approximately 1,175,000 transition-metal oxides and fluorides, followed by DFT validation of the most promising candidates. The proposed workflow enables filtering of large candidate sets and accelerates the potential discovery of low volume change intercalation materials for batteries.

\end{abstract}

\section{Introduction}\label{sec:intro}

Active electrode materials of ion-intercalation batteries often undergo structural changes during cycling, which can adversely affect the lifetime of the batteries \cite{edgeLithiumIonBattery2021,penderElectrodeDegradationLithiumIon2020,hanReviewKeyIssues2019,hausbrandFundamentalDegradationMechanisms2015}. Materials exhibiting a volume change of less than 1\% during charge–discharge cycles, in which ions are intercalated and deintercalated into and out of the structure, are often referred to as \textit{zero-strain} materials in the literature. However, even if the overall volume change is small, anisotropic changes of the lattice parameters can still lead to significant strains along different crystallographic directions, making the term \textit{zero-strain} potentially misleading. Although such anisotropic strains can result in detrimental effects similar to those mentioned above, they may be partially compensated in polycrystalline materials. Therefore, in this work we consider structures exhibiting less than 1\% volume change during charge–discharge and refer to them as \textit{low volume change} (LVC) materials. The reduced strains and stresses at interfaces in the microstructure make LVC materials suitable for long-life electrochemical storage. 

Atomic-level simulations can assist in the search for new LVC candidates, and several studies have revealed trends across various crystallographic structures. Zhao \textit{et al.} identified design principles for materials with a face-centered cubic anion framework, where the type of transition metal (TM) and the geometry of the Li intercalation site exhibited similar trends across different structures \cite{xinyezhaoDesignPrinciplesZerostrain2022}. Wang \textit{et al.} \cite{wangQuantitativeStructurepropertyRelationship2017b} calculated the volume change for 28 oxides in either spinel or layered Li$X$O$_{2}$ structures, using 34 structural descriptors in a partial least squares analysis. They found that the ionic radius of the TM and its local environment have the strongest influence on the volume change. In a related work \cite{wangFactorsThatAffect2022}, Wang \textit{et al.} unveiled correlations between bond lengths and the structural flexibility of host frameworks using first-principles calculations. Altogether, these studies demonstrate that trends of volume changes across structures can be identified and that the extent of those volume changes can be estimated without time-consuming \textit{ab initio} calculations. However, such investigations have so far been limited to specific crystal structure types.

\begin{figure}
    \centering
    \includegraphics[width=0.6\columnwidth]{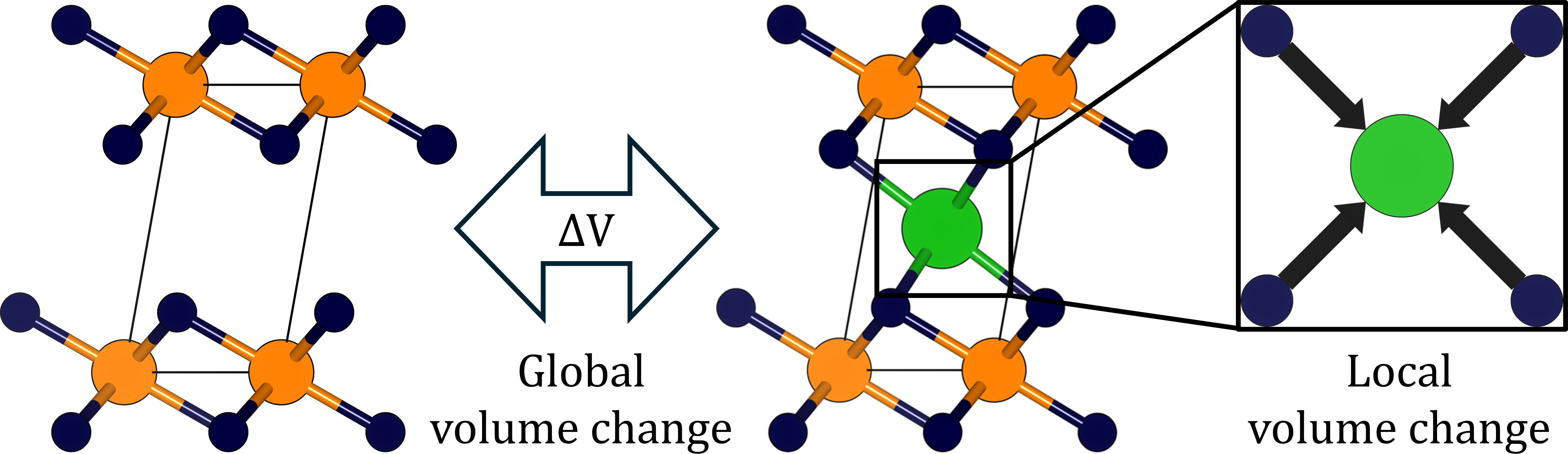}
    \caption{Schematic representation of the intercalation process: In a solid-solution mechanism, the intercalation of an atom (green sphere) into a host structure alters the local structure and can influence the global volume (i.e., the volume of the unit cell), but does not drastically modify the topology of the host lattice. The zoom-in illustrates a possible attraction of the surrounding atoms to the newly inserted atom.}
    \label{fig:mod:glob_lok_volume_changes}
\end{figure}

In this work, we present a workflow that calculates volume changes during intercalation, using bond lengths deduced by a machine learning model between all ions in a structure. The predicted bond lengths are reliable only for elements and oxidation states present in the model's training data. Within this compositional domain, the workflow can handle diverse crystallographic structures, i.e., it is not restricted to the space groups of the structures in the training dataset. The methodology is based on two main assumptions:

\begin{table*}
\caption{Selected structure pairs for which experimental volume changes ($\Delta V_{\mathrm{exp}}$) during solid-solution intercalation have been reported, along with their corresponding computed volume changes ($\Delta V_{\mathrm{DFT}}$) from the Materials Project dataset. The space group is the same for the empty and filled host structures.}
\centering
\begin{tabular}{lll
	S[table-format=2.2]
	S[table-format=2.2]}
	\hline
	Host structure ($S^{\mathrm{host}}_\mathrm{DFT}$) &
	Occupied structure ($S^{\mathrm{occ.}}_\mathrm{DFT}$) &
	Space group &
	{$\Delta V_\mathrm{DFT}$ [\%]} &
	{$\Delta V_{\mathrm{exp}}$ [\%]} \\
	\hline
	FePO$_4$               & LiFePO$_4$                  & \textit{Pnma}    & 6.98  & 6.77  \cite{zhangStructurePerformanceLiFePO42011} \\
	FePO$_4$               & NaFePO$_4$                  & \textit{Cmcm}    & 19.67 & 17.58 \cite{casas-cabanasCrystalChemistryNa2012} \\ 
	Ni$_{0.5}$Mn$_{1.5}$O$_{4}$ & Li$_{2}$Ni$_{0.5}$Mn$_{1.5}$O$_{4}$ & \textit{P4$_{3}$32} & 5.32  & 5.83  \cite{wangFactorsThatAffect2022} \\
	\hline
	\end{tabular}
	\label{tab:strukturpaare}
\end{table*}

(i) The bond length between ions primarily depends on their oxidation states and on their local environment, and it is therefore similar for a variety of different structures. This concept is related to the assignment of ionic radii, based on an ion's oxidation state and coordination number. The bond length can then be considered as the sum of the ionic radii of the two ions forming the bond. Shannon's table of ionic radii is a well-known example, in which the radii of 475 ionic-species and oxidation-state combinations were determined from a variety of experimental crystal structure data \cite{shannonRevisedEffectiveIonic1976}.

(ii) The intercalation occurs in a solid-solution manner, i.e., the crystal structure is not fundamentally altered. In some materials, the small volume change is instead caused by a structural phase transition, as in the well-known lithium titanate, where the spinel structure transforms into a rock salt structure during the insertion of Li ions \cite{ohzukuZeroStrainInsertionMaterial1995,ziebarthLithiumDiffusionSpinel2014}. For cathodes, more LVC materials are known that follow a solid-solution mechanism \cite{ariyoshiZerostrainInsertionMechanism2005,gaoZeroStrainLi2VSiO5HighPerformance2023,konumaDimensionallyInvariableHighcapacity2023,debiasiLiCaFeF6Zerostrain2017}. These materials have local structural compensation mechanisms that result in minimal volume changes and good reversibility \cite{zhaoZerostrainCathodeMaterials2022}. Komayko \textit{et al.} compared the performance of two materials with identical elemental compositions but different intercalation mechanisms. Their detailed electrochemical study revealed that the solid-solution material outperformed the one with a two-phase mechanism with respect to the retention of capacity and energy density \cite{komaykoAdvantagesSolidSolution2023}. The solid-solution intercalation process is schematically depicted in Figure~\ref{fig:mod:glob_lok_volume_changes}.

In previous studies, we have used density functional theory (DFT) to investigate several novel LVC candidate materials and identified trends that support the two assumptions made above. In one study \cite{baumannFirstprinciplesAnalysisInterplay2023} of compounds of the type Li$_x$Ca$M$F$_6$ (colquiriites), where $M$ denotes a \textit{3d}-transition-metal element, we observed a clear correlation between the change in the ionic radius of the $M$ cation with increasing $x$ and the corresponding change in the volume of the surrounding fluorine-anion octahedron. We also found that the local coordination environments around each cation in the structure can influence the global volume change, independent of whether the cation is intercalated or electrochemically active. In another study \cite{baumannFirstprinciplesStudyStrain2024a}, we calculated the volume changes associated with the intercalation of Li$^+$, Na$^+$, and K$^+$ into three distinct tungsten–bronze–type structures, revealing a linear relationship between the ionic radius of the intercalated ion and the local volume change at the intercalation site. Although DFT calculations can give results in good agreement with experimental results (see Table~\ref{tab:strukturpaare}), their computational complexity and substantial expenditure of time limit their suitability for large-scale screening of new materials. The machine learning model underlying this workflow was trained and tested on a dataset of \textit{3d} TM oxides and fluorides obtained from DFT calculations. Within this compositional domain and for various structure types, the workflow serves as an efficient pre-screening tool to prioritize promising candidates from sets of material data.

The paper is organized as follows: In Sec.\ \ref{sec:theory}, we first describe the used dataset (Sec.\ \ref{sec:theory:dataset}), then introduce the atomic descriptors used to deduce bond lengths (Sec.\ \ref{sec:theory:lsop}), and present the surrogate model that predicts the volume change using those bond lengths (Sec.\ \ref{sec:mod:theory:vol}). In Sec.\ \ref{sec:res:bond_lengths}, we present the results of the deduced bond lengths and compare them to pair-wise sums of the ionic radii instead. The performance of the workflow is reported and discussed in Sec.\ \ref{sec:res:volume}. In Sec.\ \ref{sec:screen}, we first describe the screening approach for new intercalation compounds (Sec.\ \ref{sec:screen:approach}), and then we present the most promising candidates validated by DFT (Sec.\ \ref{sec:screen:disc}). Finally, a summary is given in Sec.\ \ref{sec:mod:conclusion}. In the supporting information (SI), we provide the computational details of the workflow, followed by a detailed description of the volume prediction model, and an illustrative application to the system FePO$_4 \leftrightarrow$ LiFePO$_4$.

\section{Methods}\label{sec:theory}
	
\begin{figure*}[h]
    \centering
    \includegraphics[width=0.75\linewidth]{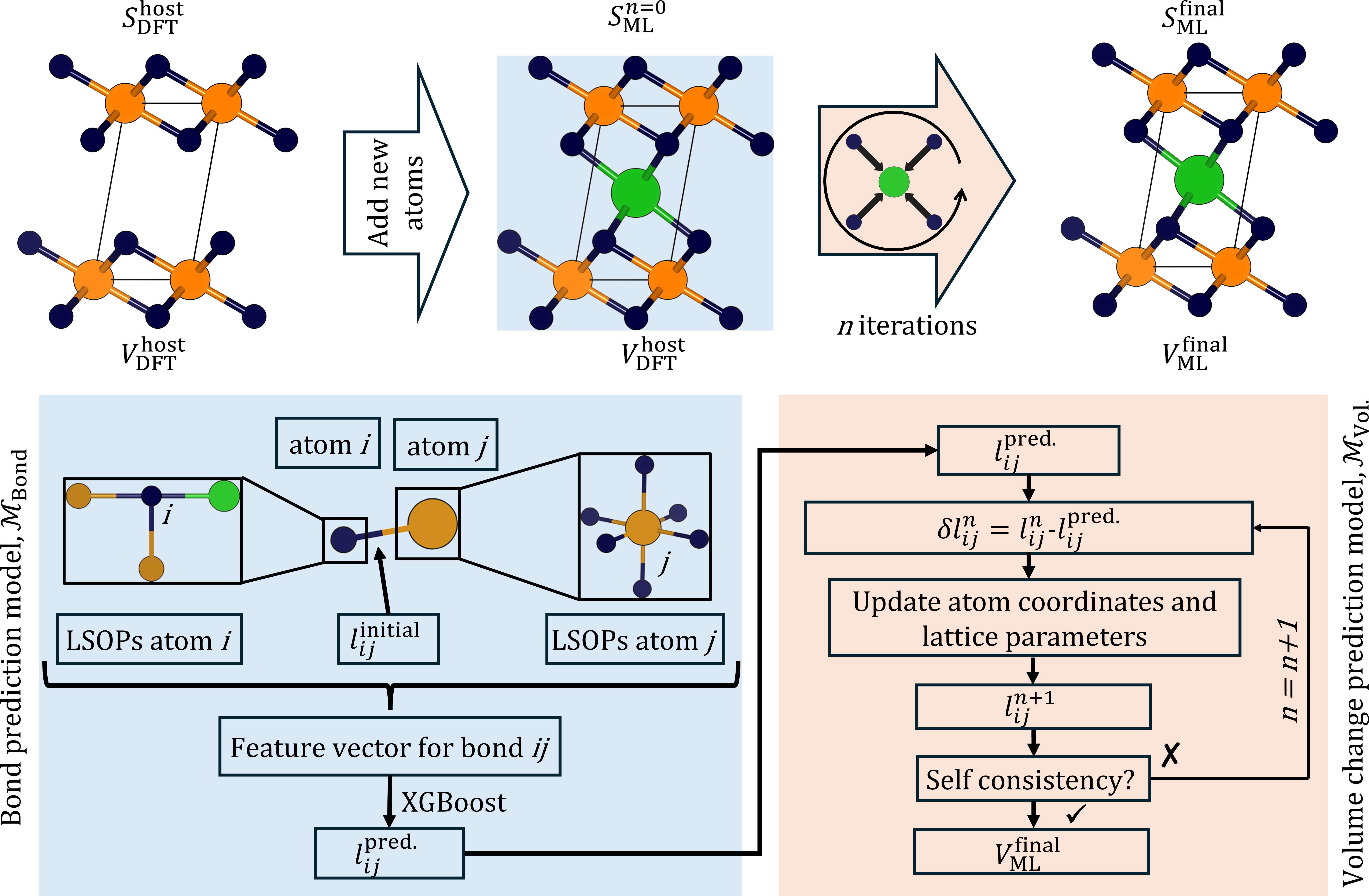}
    \caption{Workflow to predict the volume change upon intercalation of ions into a host structure ($S^{\mathrm{host}}_\mathrm{DFT}$). In the first step, this results in a structure $S^{n=0}_\mathrm{ML}$ with the unchanged volume $V^{\mathrm{host}}_\mathrm{DFT}$. For this structure, atomic and bond feature vectors are created to predict the bond lengths ($l^{\mathrm{pred.}}_{ij}$). By systematically shifting atomic positions and cell vectors, the structure is then iteratively adjusted to minimize the differences between the actual bond lengths $l^{\mathrm{actual}}_{ij}$ and the predicted values $l^{\mathrm{pred.}}_{ij}$ based on the model $\mathcal{M}_{\mathrm{Bond}}$. The iteration is continued until self-consistency is achieved, yielding the final structure $S^{\mathrm{final}}_\mathrm{ML}$ with volume $V^{\mathrm{final}}_\mathrm{ML}$. The procedures highlighted in light blue and red are described in detail in Sec. \ref{sec:theory:lsop} and \ref{sec:mod:theory:vol}, respectively.}
    \label{fig:whole_workflow}
\end{figure*}

This section describes the methods employed in the workflow to predict volume changes, with the complete procedure depicted in Figure \ref{fig:whole_workflow}. The workflow consists of two models: a model for predicting bond lengths (denoted as $\mathcal{M}_{\mathrm{Bond}}$ in the following) and a model for predicting volume changes ($\mathcal{M}_{\mathrm{Vol.}}$). Section \ref{sec:theory:dataset} presents the dataset used to train and test the workflow. Section \ref{sec:theory:lsop} first introduces the local structure order parameters (LSOPs), which characterize the local geometric environment around an atom. The LSOPs are used in $\mathcal{M}_{\mathrm{Bond}}$ to obtain bond lengths between neighboring atoms. Section \ref{sec:mod:theory:vol} presents the surrogate model $\mathcal{M}_{\mathrm{Vol.}}$ for calculating volume changes using these bond lengths.

\subsection{DFT dataset}\label{sec:theory:dataset}

We used a dataset of DFT results provided by the \textit{Materials Project} (MP) \cite{jainCommentaryMaterialsProject2013}. This dataset contains 5602 structure pairs, each consisting of one structure at low or zero concentration of the intercalated cation ($S^{\mathrm{host}}_\mathrm{DFT}$) and another one at a higher concentration of that cation ($S^{\mathrm{occ.}}_\mathrm{DFT}$). We define the volume change $\Delta V_\mathrm{DFT}$ as:

\begin{equation}
     \Delta V_\mathrm{DFT} = \frac{V_\mathrm{DFT}^\mathrm{final} - V_\mathrm{DFT}^\mathrm{host}}{V_\mathrm{DFT}^\mathrm{host}} \times 100\%,
     \label{eq:vol_change}
\end{equation}
where $V_\mathrm{DFT}^\mathrm{host}$ and $V_\mathrm{DFT}^\mathrm{final}$ are the volumes of the structures $S^{\mathrm{host}}_\mathrm{DFT}$ and $S^{\mathrm{occ.}}_\mathrm{DFT}$, respectively. From the dataset, we selected all \textit{3d}-TM oxides and fluorides for which oxidation states could be assigned using the methods of MP (the detailed procedure is described in section S1 of the SI). Oxides were included because they are numerous and well studied, providing a solid experimental and theoretical foundation \cite{campeonFundamentalsMetalOxide2021}. Fluorides, on the other hand, may enable higher operating voltages and are therefore particularly promising for new electrode materials to be discovered by screening studies \cite{koyamaNewFluorideCathodes2000,lemoineFluorinatedMaterialsPositive2022}. We further restricted the selection to structures with an absolute volume change of less than 30\% in the DFT calculations. This threshold was applied according to the assumption of a solid-solution insertion mechanism. For model training, the input structures were generated by taking the host structure without intercalated cations and then inserting the cations at the same crystallographic positions as in the corresponding intercalated structure. Structural correspondence between the constructed occupied and reference intercalated structures was subsequently verified using an explicit structure-matching procedure implemented in the Python package \textit{pymatgen} \cite{ongPythonMaterialsGenomics2013}. Three representative examples are listed in Table~\ref{tab:strukturpaare}.

\subsection{Model for predicting bond lengths, $\mathcal{M}_{\mathrm{Bond}}$}\label{sec:theory:lsop}

\begin{figure}[h]
\centering
\includegraphics[width=0.7\columnwidth]{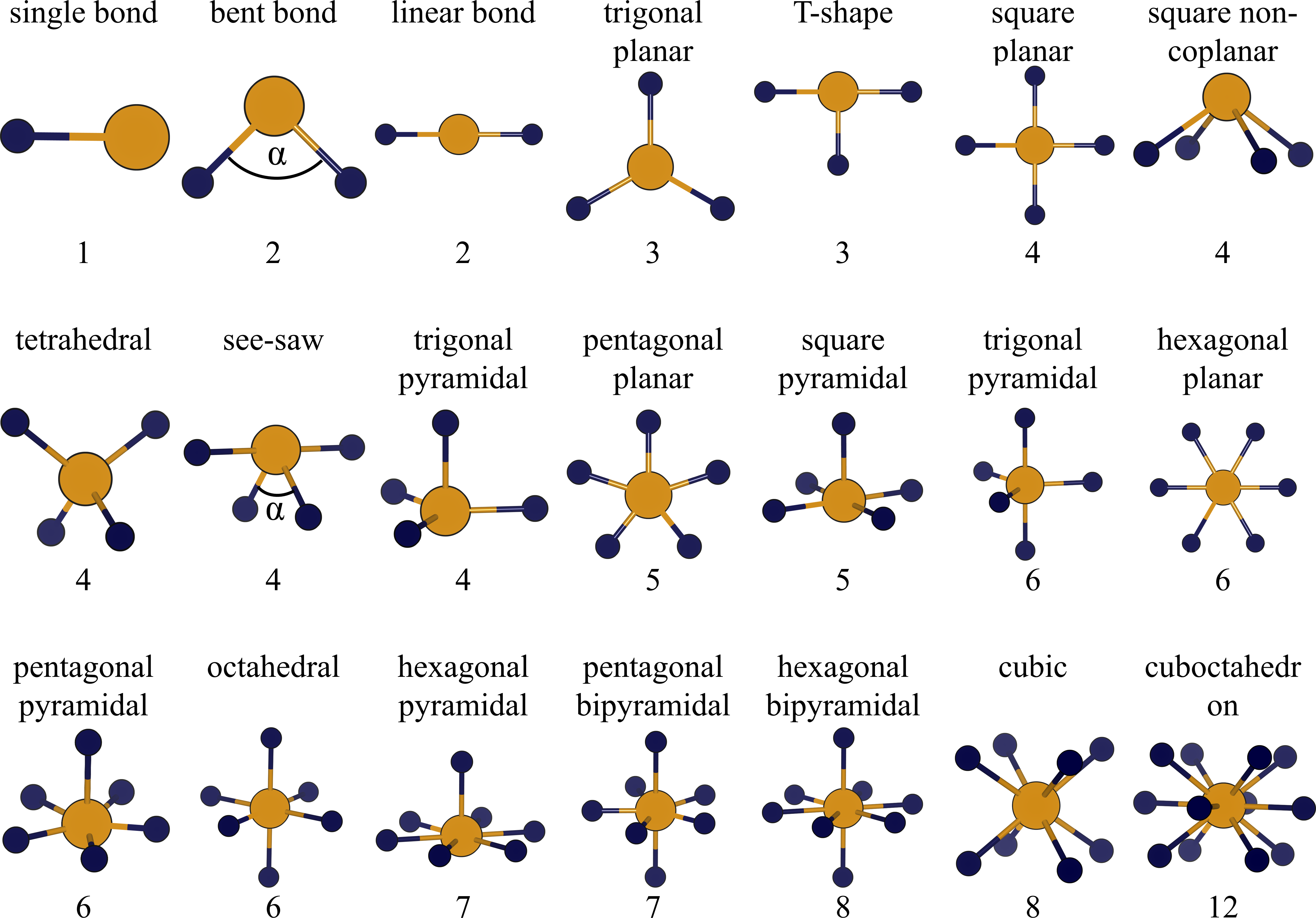}
\caption{LSOP structure motifs used in this work. The motifs describe the coordination of the central atom (orange) by the neighboring atoms (dark blue). The values 90$^\circ$, 104.5$^\circ$, 120$^\circ$ and 150$^\circ$ for $\alpha$ were used for the bent bond and the values 90$^\circ$ and 120$^\circ$ for $\alpha$ were used for the see-saw shaped coordination leading to 25 LSOPs in total. The coordination number, by which the similarity to a motif is multiplied, is given below each motif and corresponds to the number of dark blue neighboring atoms. The Figure was adapted from Fig.\ 6 in [N.\ E.\ R.\ Zimmermann and  A.\ Jain, RSC Adv. 2020, 10, 6063--6081]. Available under a CC-BY 3.0 license. Copyright N.\ E.\ R.\ Zimmermann and  A.\ Jain}
\label{fig:lsops}
\end{figure}

Similar to estimating the bond length as the sum of the two ionic radii of the bonded ions, our approach uses atomic descriptors to predict the bond length between two atoms. Baloch \textit{et al.} tabulated ionic radii for 987 combinations of element, oxidation state, and coordination number \cite{balochExtendingShannonsIonic2021a}. In contrast, our method employs the LSOPs introduced in Ref.~\citenum{zimmermannLocalStructureOrder2020a} to describe the local atomic environment. LSOPs quantify the similarity of an atom’s coordination environment to predefined structural motifs (e.g., to an ideal octahedron). The 25 LSOP motifs used in this work are illustrated in Fig.~\ref{fig:lsops}.

Each central atom is assigned a continous value between 0 and 1 for each LSOP motif, where 0 represents no similarity and 1 denotes equality. For example, in the case of an octahedron, a value of 1 corresponds to all bond angles between the central atom and its coordinating atoms being exactly 90$^\circ$. This motif similarity is further weighted by the similarity to the corresponding coordination number, also ranging from 0 to 1. In the case of an ideal octahedron, a value of 1 would correspond to equal distances to six neighboring atoms. This weighting ensures that motifs with the correct coordination number are emphasized: while an octahedron contains sub-motifs such as single bonds, angular bonds (90$^\circ$), linear bonds, or T-shapes, these receive low scores due to their mismatch in coordination number.

In total, using coordination numbers from 1 to 24, we derive 49 geometric features per atom. Considering as additional features the atomic number and the oxidation state, each atom is thus represented by a feature vector of length 51. Two atoms forming a bond are accordingly defined by a bond feature vector of length 103, where the one further component is the length of the bond in the initial structure ($S^{n=0}_\mathrm{ML}$). These feature vectors serve as input to a gradient boosting regression model to predict the final bond lengths (of the structure $S^{\mathrm{occ.}}_\mathrm{DFT}$), as implemented in the Python package \textit{XGBoost} \cite{chenXGBoostScalableTree2016}. Gradient boosting models use decision trees \cite{breimanClassificationRegressionTrees2017} as base learners, enabling them to capture complex non-linear relationships, to handle heterogeneous feature types, and to deliver high predictive accuracy with efficient training times \cite{friedmanGreedyFunctionApproximation2001}. In Sec.\ \ref{sec:res:bond_lengths}, we compare the bond lengths predicted by $\mathcal{M}_{\mathrm{Bond}}$ with those obtained from the ionic radii tabulated by Baloch \textit{et al.} \cite{balochExtendingShannonsIonic2021a}. Computational details of the ML tools and settings are given in section S1 of the SI.

\subsection{Model for predicting volume changes, $\mathcal{M}_{\mathrm{Vol.}}$}\label{sec:mod:theory:vol}

In the following, we present the surrogate model, $\mathcal{M}_{\mathrm{Vol.}}$, to calculate the volume change upon intercalation. Structures and results predicted with $\mathcal{M}_{\mathrm{Vol.}}$ carry the subscript ML (machine learning), as they are typically derived using the bond lengths predicted by $\mathcal{M}_{\mathrm{Bond}}$ as input. Importantly, $\mathcal{M}_{\mathrm{Vol.}}$ itself contains no learnable parameters; given an input structure and a set of bond lengths, it performs a fixed computation without any training or parameter tuning.

The procedure begins with a host structure $S^{\mathrm{host}}_\mathrm{DFT}$ into which additional cations are inserted, resulting in the initial, non-equilibrium structure $S^{n=0}_\mathrm{ML}$. The workflow was first evaluated with the dataset described in Sec.\ \ref{sec:theory:dataset}, i.e., the additional atoms were inserted at the fractional coordinates as in the corresponding occupied structures $S^{\mathrm{occ.}}_\mathrm{DFT}$. Atomic feature vectors are then determined for all atoms in $S^{n=0}_\mathrm{ML}$, and the bond lengths between neighboring atoms $i$ and $j$, which the intercalated structure should exhibit in equilibrium (i.e., in the structures $S^{\mathrm{occ.}}_\mathrm{DFT}$), are predicted by the model $\mathcal{M}_{\mathrm{Bond}}$ (Sec. \ref{sec:theory:lsop}). The obtained bond lengths, $l^{\mathrm{pred.}}_{ij}$, generally differ from the actual bond lengths $l^{n=0}_{ij}$ in $S^{n=0}_\mathrm{ML}$, for example due to changes in oxidation-state or lattice relaxations. The difference between $l^{\mathrm{pred.}}_{ij}$ and $l^{n=0}_{ij}$ is used to adjust the structure (atomic positions and lattice parameters), leading, in the first step, to a structure $S^{n=1}_\mathrm{ML}$. The details of this procedure are presented in section S2 of the SI. The volume of the updated structure $S^{n=1}_\mathrm{ML}$ is calculated from its lattice parameters and compared to the volume $S^{n=0}_\mathrm{ML}$. If the difference is above a predefined threshold, the resulting bond lengths $l^{n=1}_{ij}$ are then compared again to $l^{\mathrm{pred.}}_{ij}$, followed by another adjustment. Repeating this procedure iteratively minimizes the differences between predicted and actual bond lengths. When self-consistency is achieved, i.e., the difference in volume between two consecutive steps falls below the threshold, the procedure terminates. The total volume change $\Delta V_\mathrm{ML}$, defined analogously to $\Delta V_\mathrm{DFT}$ in Eq.~(\ref{eq:vol_change}), is then computed between the final structure $S^{\mathrm{final}}_\mathrm{ML}$ and the initial structure $S^{n=0}_\mathrm{ML}$. For the structure pairs in the dataset, this model-predicted volume change can then be directly compared to the DFT-calculated volume change between $S^{\mathrm{occ.}}_\mathrm{DFT}$ and $S^{\mathrm{host}}_\mathrm{DFT}$ to assess the workflow's performance. In section S3 of the SI we illustrate the iterative procedure for a representative example structure.

\section{Results and Discussion}

\subsection{Performance of the workflow}\label{sec:res}

In this section, we first present the results obtained with the model $\mathcal{M}_{\mathrm{Bond}}$ described in Sec.\ \ref{sec:theory:lsop} and compare them to bond lengths based on the sum of the ionic radii. The better agreement of our approach to the DFT derived bond lengths is also reflected in the improved results for the volume changes, which are presented in Sec.\ \ref{sec:res:volume}.

\subsubsection{Prediction of bond lengths}\label{sec:res:bond_lengths}

\begin{figure}
\centering
\includegraphics[width=0.7\columnwidth]{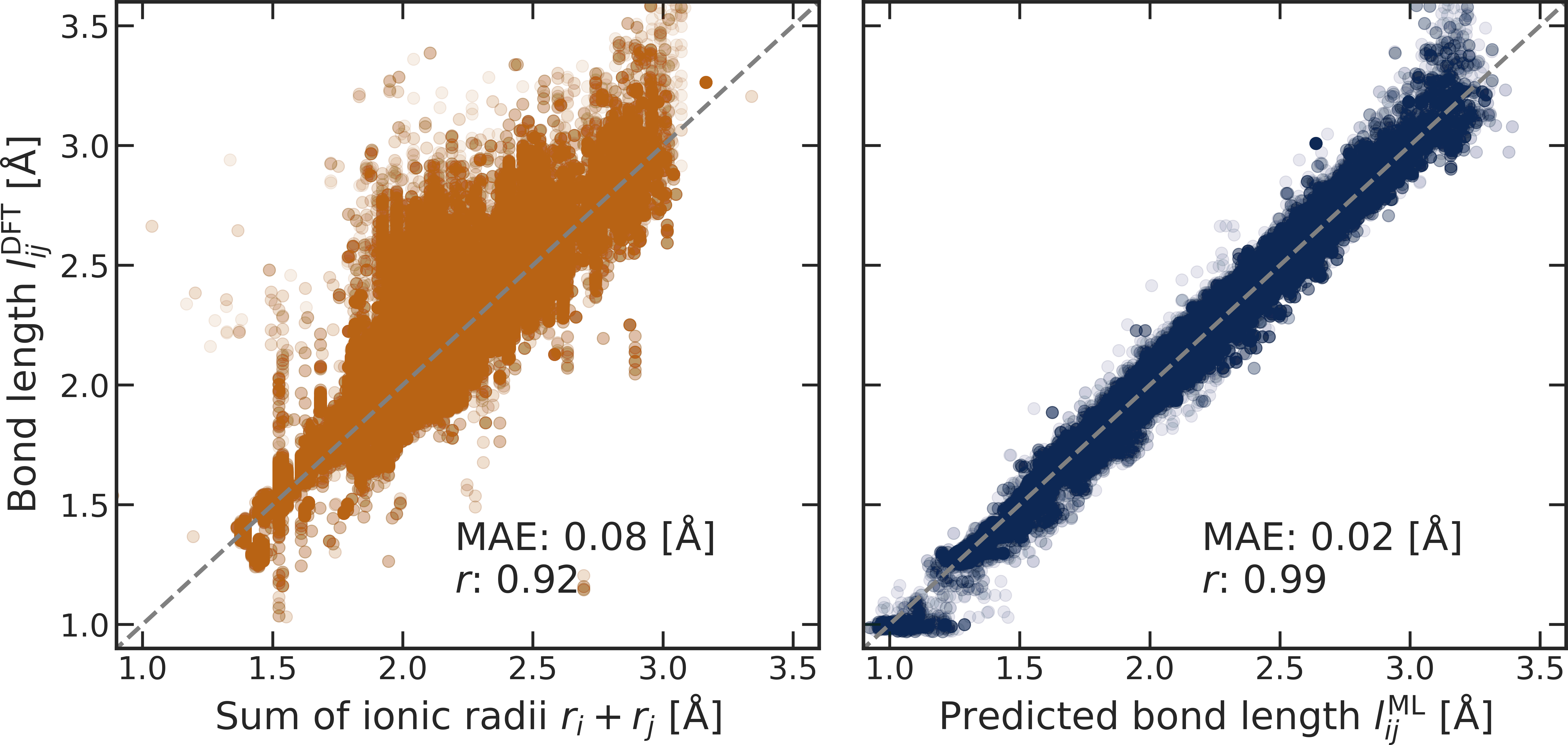	}
\caption{Left: The DFT-derived bond lengths between two atoms $i$ and $j$ plotted against the sum of their ionic radii. Right: The DFT-derived bond lengths plotted against the obtained bond lengths using the ML model $\mathcal{M}_{\mathrm{Bond}}$. The MAE and the Pearson correlation coefficient ($r$) for both datasets are given in the respective plots. Each data point is shown with an opacity of 0.01, so that 100 overlapping points result in full coverage.}
\label{fig:bond_lengths}
\end{figure}

The first assumption made in Sec.\ \ref{sec:intro} was that pairs of ions (defined by their elemental types and oxidation states) in similar local environments have similar bond lengths in different crystal structures. One way to express this is through ionic radii. In this concept, ions are considered as spheres that have similar radii in different environments. The bond length is then simply the sum of the radii of the two ions forming the bond. For all structures in the dataset, Figure \ref{fig:bond_lengths} (left graph) shows the sum of the ionic radii for all bonds between neighboring ions compared to the bond length calculated by DFT. The ionic radii were taken from the database of Baloch \textit{et al.} \cite{balochExtendingShannonsIonic2021a}, which contains nearly twice as many tabulated values as Shannon's original work \cite{shannonRevisedEffectiveIonic1976}. For ions with non-integer oxidation states, we used an arithmetic mean value.

There is a clear correlation between the sum of ionic radii and the bond length. This is reflected in the value of the Pearson correlation coefficient, $r$, which quantifies the linear correlation between two variables. When comparing ML and DFT values for the bond lengths, higher values of $r$ indicate better predictive performance \cite{pearsonVIINoteRegression1997}. While the mean absolute error (MAE) is below 0.1~\AA, the plot also shows much larger deviations for some bond lengths. These deviations can have an adverse impact on the volume prediction (see Sec.\ \ref{sec:res:volume}).

In contrast to using the sum of the ionic radii to deduce the bond length, which incorporates only the coordination number as information about the local environment, the ML model $\mathcal{M}_{\mathrm{Bond}}$ uses a total of 51 descriptors per ion, 49 of which are directly related to the local structure. The right graph of Figure \ref{fig:bond_lengths}\ displays the model's predictions as derived from a 5-fold cross-validation. The cross-validation split was performed on structure pairs, such that each host structure and its corresponding intercalated structure were always assigned to the same subset, preventing information leakage between training and validation datasets. Additionally, in each split, the space group of a structure appeared exclusively in either the training or validation set, enabling an evaluation of the model’s generalization across different crystal structures. It was ensured that each 3\textit{d} transition metal appeared at least once in the training set, so that bond length predictions remain meaningful and the evaluation focuses on transferability across structures rather than across different elements. Only the bond-length predictions made on the respective validation sets are shown, i.e., the model had no prior exposure to those data points during the training. The resulting low MAE and high Pearson correlation coefficient $r$ reflect the stronger predictive performance of the model $\mathcal{M}_{\mathrm{Bond}}$, compared to the sum of ionic radii.

\subsubsection{Prediction of volume changes}\label{sec:res:volume}

To validate the volume surrogate model $\mathcal{M}_{\mathrm{Vol.}}$, we applied the procedure described in Sec.\ \ref{sec:mod:theory:vol} to the structure pairs in the dataset (see Sec.\ \ref{sec:theory:dataset}). That is, we employed the DFT-derived bond lengths of the structures $S^{\mathrm{occ.}}_\mathrm{DFT}$ for the terms $l^{\mathrm{pred.}}_{ij}$ [see Eq.\ (S1)]. Using these bond lengths as an upper-bound check, the resulting structures $S^{\mathrm{final}}_\mathrm{ML}$ are expected to closely resemble $S^{\mathrm{occ.}}_\mathrm{DFT}$, and the predicted volume changes should closely match the volume differences between $S^{\mathrm{occ.}}_\mathrm{DFT}$ and $S^{\mathrm{host}}_\mathrm{DFT}$. The validation yields a median absolute error of approximately 0.004\% and a 99th-percentile absolute error of 0.12\% (i.e., 99\% of the deviations are smaller than 0.12\%). This low deviation confirms that $\mathcal{M}_{\mathrm{Vol.}}$ accurately reproduces DFT-calculated volume changes and validates the underlying model assumptions.

\begin{table}[H]
\caption{Metrics for the volume changes predicted by the model $\mathcal{M}_{\mathrm{Vol.}}$ using different bond length inputs. All results are compared to the DFT volume changes, $\Delta V_\mathrm{DFT}$. Precision and recall for identifying structure pairs with $|\Delta V_\mathrm{DFT}| < 1\%$ at two thresholds are also provided. For $\mathcal{M}_{\mathrm{Bond}}$, the reported values correspond to the averages over the test sets of a 5-fold cross-validation.}
\centering
\begin{tabular}{llcc}
\hline
\multicolumn{2}{l}{Metric} & Ionic radii sum $r_i+r_j$ & $\mathcal{M}_{\mathrm{Bond}}$ $l^{\mathrm{ML}}_{ij}$ \\
\hline
\multicolumn{2}{l}{Mean signed error [\%]} & -6.43 & -0.24 \\
\multicolumn{2}{l}{Median absolute error [\%]} & 6.02 & 2.04 \\
\multicolumn{2}{l}{80th-percentile absolute error [\%]} & 9.78 & 4.68 \\
\multicolumn{2}{l}{90th-percentile absolute error [\%]} & 12.82 & 6.86 \\
\multicolumn{2}{l}{95th-percentile absolute error [\%]} & 16.26 & 8.96 \\
\multirow{2}{*}{Threshold $|V_{\mathrm{ML}}|$ < 1\%} & Precision & 0.08 & 0.37\\
& Recall & 0.05 & 0.29\\
\multirow{2}{*}{Threshold $|V_{\mathrm{ML}}|$ < 2\%} & Precision & 0.09 & 0.32\\
& Recall & 0.12 & 0.54\\
\hline
\end{tabular}
\label{tab:metrics}
\end{table}

As shown in Figure~\ref{fig:bond_lengths}, estimating bond lengths from the sum of the ionic radii already provides a reasonably good approximation despite its simplicity, typically achieving an accuracy better than 0.1~\AA. However, this level of accuracy is insufficient for reliable volume predictions when these bond lengths are used as input for $\mathcal{M}_{\mathrm{Vol.}}$, as reflected by the performance metrics summarized in Table~\ref{tab:metrics}. In particular, the mean signed error of -6.43\% indicates a systematic underestimation of the volume changes. Moreover, the median absolute error of 6.02\% implies that 50\% of the predicted volume changes deviate by more than 6\% from the corresponding $\Delta V_\mathrm{DFT}$ values.

To further assess the practical utility of the model, we evaluated its performance in terms of precision and recall for different thresholds used to classify structure pairs as promising candidates with low volume changes \cite{saitoPrecisionRecallPlotMore2015}. Here, an actual positive is defined as a structure pair with $|\Delta V_\mathrm{DFT}| < 1\%$. A predicted positive is defined as a structure pair for which the absolute volume change predicted by $\mathcal{M}_{\mathrm{Vol.}}$ falls below a chosen threshold. Precision is then defined as the fraction of predicted low-volume-change structure pairs that are indeed low-volume-change according to DFT, that is, the number of true positives divided by the total number of predicted positives (true positives plus false positives). It therefore quantifies how reliable the model predictions are when identifying promising candidates. Recall is defined as the fraction of all truly low-volume-change structure pairs that are correctly identified by the model, that is, the number of true positives divided by the total number of actual positives (true positives plus false negatives). It thus measures how completely the model captures the relevant candidates. The higher both values are, the better the performance of the model.

However, when using bond lengths derived from the sum of the ionic radii, both metrics remain unsatisfactory. For example, at a threshold of $|V_{\mathrm{pred.}}| < 2\%$, only 9\% of the predicted low-volume-change structure pairs are actually low-volume-change pairs (precision = 9\%), and this threshold captures only 12\% of all true low-volume-change structure pairs (recall = 12\%). Such performance is clearly inadequate for large-scale screening applications. A notable limitation is the treatment of non-integer oxidation states, which requires assigning intermediate values. Moreover, to rely solely on the coordination number as a local structure descriptor is apparently too crude.

\begin{figure}
\centering
\includegraphics[width=0.5\columnwidth]{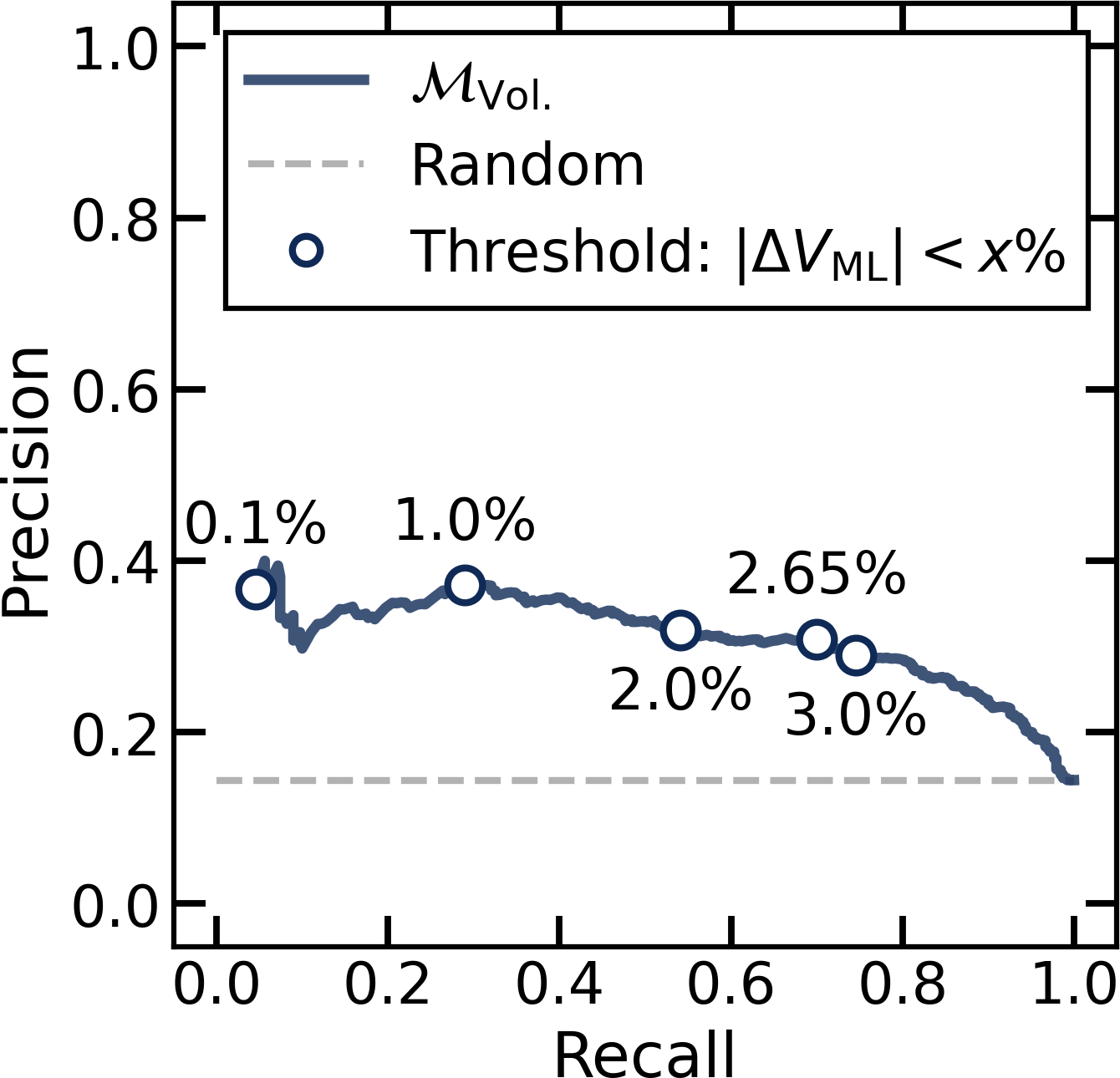}
\caption{Precision–-recall curve for the model $\mathcal{M}_{\mathrm{Vol.}}$, evaluated using bond lengths predicted by the model $\mathcal{M}_{\mathrm{Bond}}$. Selected thresholds of $|\Delta V_{\mathrm{ML}}|$ are indicated with circles. A baseline corresponding to random selection of structure pairs is shown.}
\label{fig:disc:prec_recall}
\end{figure}

Next, we tested the model $\mathcal{M}_{\mathrm{Vol.}}$ for the structure pairs in the dataset by now taking the values derived by the bond prediction model $\mathcal{M}_{\mathrm{Bond}}$ for the bond lengths $l^{\mathrm{pred.}}_{ij}$ of the intercalated structures (see Sec.\ \ref{sec:theory:lsop} and Fig.\ \ref{fig:bond_lengths}). Because this setting corresponds to predictions for structures unseen during training, it indicates the situation encountered in high-throughput screening of new candidates beyond the dataset (Sec.\ \ref{sec:screen}). We describe this procedure in detail in section S3 of the SI for the example of the structure pair FePO$_4 \leftrightarrow$ LiFePO$_4$. Compared to the approach based on the sum of ionic radii, this method exhibits a substantially smaller mean signed error of $-0.24\%$, indicating the absence of a pronounced systematic under- or overestimation. Furthermore, the median absolute error and the corresponding percentiles are significantly reduced, with 80\% of the structure pairs showing a deviation from $\Delta V_\mathrm{DFT}$ below 5\%. The precision and recall values also improve when using $\mathcal{M}_{\mathrm{Bond}}$ to predict the bond lengths $l^{\mathrm{pred.}}_{ij}$ instead of relying on the sum of ionic radii. At a threshold of 2\%, more than 50\% of the structure pairs with $|\Delta V_\mathrm{DFT}| < 1\%$ are identified. The corresponding precision--recall curve is shown in Fig.~\ref{fig:disc:prec_recall}, where selected thresholds are highlighted and the performance is compared to a baseline in which structure pairs are selected at random.

To determine a suitable threshold for screening, we employed the F$_1$ score \cite{rijsbergenInformationRetrieval1979}, which provides a single metric that balances precision and recall. This choice reflects the need to identify as many LVC structures as possible while keeping the computational cost of validation calculations manageable. The maximum F$_1$ score of 0.43 is obtained at a threshold of $|\Delta V_\mathrm{ML}| < 2.65\%$. At this threshold, approximately 70\% of the structures with $|\Delta V_\mathrm{DFT}| < 1\%$ are successfully identified (recall = 0.70), while 31\% of the predicted structure pairs indeed satisfy $|\Delta V_\mathrm{DFT}| < 1\%$ (precision = 0.31). One reason for this moderate precision may be that $\mathcal{M}_{\mathrm{Bond}}$ considers only the atoms involved in the bond as input features. In reality, however, bond lengths are influenced by a complex interplay between local and global structural environments. Moreover, as shown in Ref.~\citenum{baumannFirstprinciplesAnalysisInterplay2023}, the magnetic configuration of the TM element also affects both, local structural features and the global structure, an effect that is not yet captured by the model $\mathcal{M}_{\mathrm{Bond}}$. Nevertheless, the improved predictive performance relative to both random selection and the approach based on the sum of ionic radii, demonstrates that the workflow already provides valuable insight and holds clear potential for further refinement.

When using the DFT bond lengths as input for $\mathcal{M}_{\mathrm{Vol.}}$, the volume prediction is highly accurate. The structure pair with the largest deviation of 0.7\% is NiO$_{2} \leftrightarrow$ Li$_{2}$NiO$_{2}$ in the space group \textit{R$\bar{3}$m}. Here, the structure undergoes strong changes in cell geometry, with the cell angles shifting from 79.7$^\circ$, 78.7$^\circ$, and 60$^\circ$ to 76.7$^\circ$, 90$^\circ$, and 60$^\circ$, respectively. The model does not account for angle variations, which partially explains this result. However, this example represents not only the largest deviation observed, but also one of only two structures in the dataset (2707 structure pairs) where bond angle changes exceeding 0.5$^\circ$, occur. As most values obtained using the DFT bond lengths achieve an accuracy within a few tenths of a percent, this indicates that the angle constraint introduces noticeable errors only in very few cases. The deviations discussed above are therefore much more likely to be attributed to inaccuracies in the bond lengths used as input for $\mathcal{M}_{\mathrm{Vol.}}$, rather than to an inherent limitation of the model itself.

\subsection{Screening of intercalation materials}\label{sec:screen}

In the following, we present a screening study of intercalation compounds using the workflow described in Sec.\ \ref{sec:theory}. We describe the generation of possible intercalation compounds and present the predicted volume changes obtained with the workflow in Sec.\ \ref{sec:screen:approach}. The most promising candidates are subsequently validated by DFT, and Sec.\ \ref{sec:screen:disc} discusses both the screening methodology and presents the most promising LVC structure pairs.

\subsubsection{Screening approach}\label{sec:screen:approach}

We considered all binary and ternary \textit{3d}-TM oxide and fluoride crystal structures from the MP database \cite{jainCommentaryMaterialsProject2013} as potential host structures $S^{\mathrm{host}}_\mathrm{DFT}$. As a first step, we filtered out structures with an energy above the convex hull exceeding 0.01 eV per atom, retaining 2108 potential host structures. This threshold ensures that only thermodynamically stable or slightly unstable structures are included, which increases the likelihood for experimental synthesis. We then excluded any remaining structures containing at least one element from the set of toxic or radioactive elements, namely, Be, As, Se, Tc, Cd, Sb, Ba, Hg, Tl, Pb, Po, At, Rn, Fr, Ra, Ac, Th, Pa, U, Np, Pu, Am, Cm, Bk, Cf, Es, Fm, Md, No and Lr, which removed 327 structures. We used the Python package \textit{pyxtal} \cite{fredericksPyXtalPythonLibrary2021} to obtain the vacant Wyckoff positions of the space group to which each host crystal structure belongs and occupied these sites with either Li, Na, K, Mg, or Ca. For Wyckoff positions with variable coordinates, e.g., $(0, y, 0)$, the variable coordinates were systematically sampled on a grid with a maximum spacing of 1~\AA along each degree of freedom, and for positions with multiple variable coordinates, all combinations were sampled. From the structures generated in this way, those containing at least one interatomic distance between cations shorter than 1.3~\AA\ were discarded.

\begin{figure}
\centering
\includegraphics[width=0.7\columnwidth]{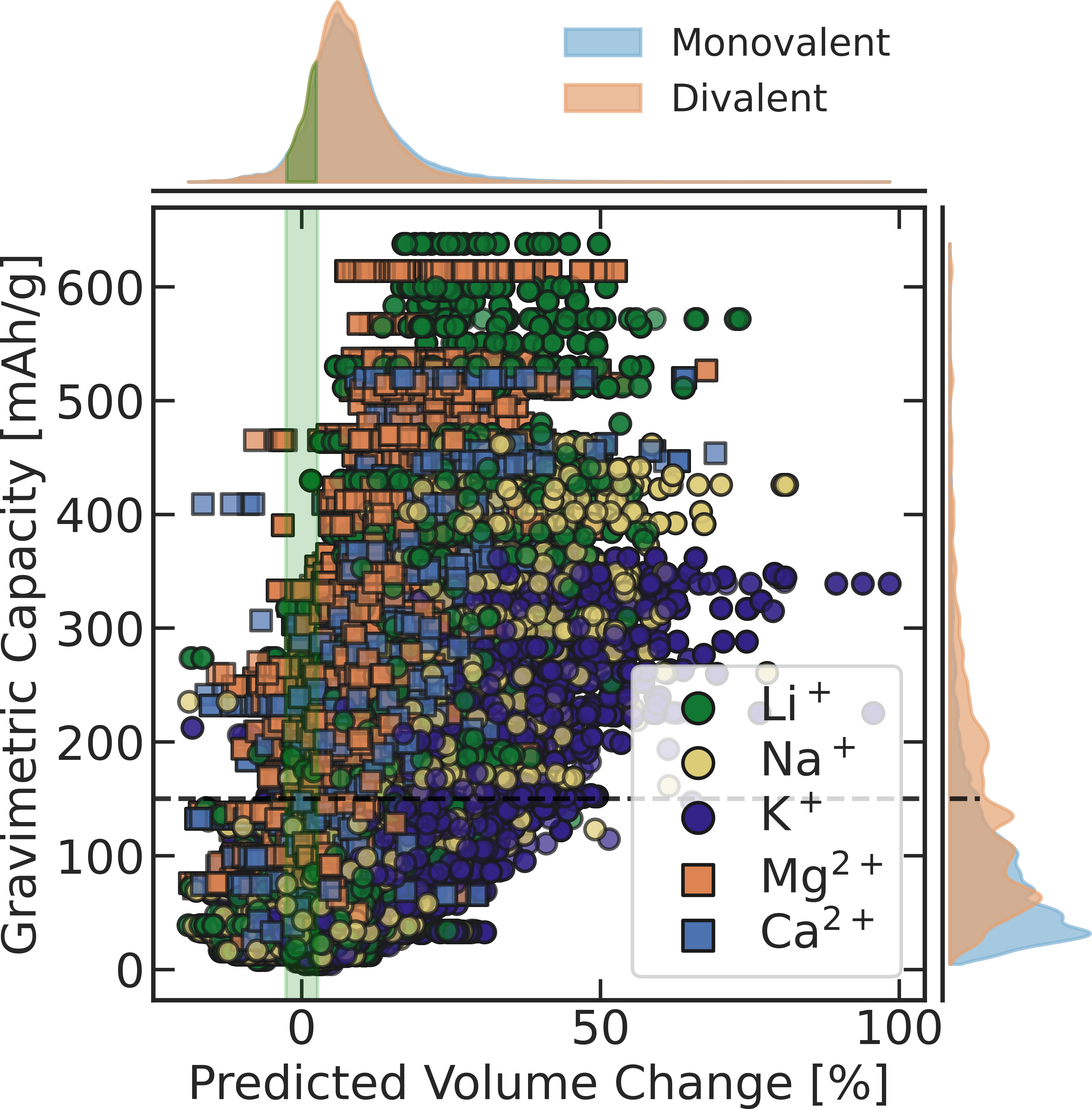}
\caption{Predicted volume changes and theoretical capacites for a total of 1,174,384 structure pairs. The data points are color labeled by the species of intercalated cation. The range of predicted low volume changes of $|\Delta V| < 2.65\%$ is highlighted by the light green bar. Out of 1,174,384 structure pairs, 168,749 (14.4\%) fall within this range. A total of 248,711 (21.2\%) structure pairs have a theoretical capacity exceeding 150 mAh/g, of which 5824 fall into volume-change range of $|\Delta V| < 2.65\%$ (and 242,887 structure pairs outside this range). Density distributions for both axes are shown separately for monovalent and divalent cations.}
\label{fig:disc:screen}
\end{figure}

Before conducting the volume-change calculations, we filtered out structures containing elements or oxidation states not present in the training dataset. For example, Cr-containing structures in which the oxidation-state assignment gave Cr($+\mathrm{I}$) were rejected because this oxidation state was not present in any structure in the training set, and bond-length predictions are therefore not expected to be reliable. Beneficially, this filtering approach removed unrealistic structures such as Cr$_6$B$_6$O$_{18} \leftrightarrow$ Li$_{36}$Cr$_6$B$_6$O$_{18}$ (theoretical capacity of 1054.81 mAh/g), where boron would need to adopt an oxidation state of $-\mathrm{III}$. This state is not observed in any known oxide compound \cite{Wiberg+2008} and was therefore also not present in our training data. Applying those constraints, the procedure resulted in 1,174,384 intercalated structures $S^{n=0}_\mathrm{ML}$, for which we calculated the atomic and bond feature vectors. The bond lengths were predicted with the model $\mathcal{M}_{\mathrm{Bond}}$ trained on all 2707 structure pairs of the MP dataset as described in Sec. \ref{sec:theory:dataset}. The predicted bond lengths of the intercalated structures were finally used to calculate the volume changes between them and their host structures with the model $\mathcal{M}_{\mathrm{Vol.}}$.

With the number of intercalated cations and the atomic masses, it is also possible to calculate the theoretical gravimetric capacity for the considered materials according to:

\begin{equation}
     Q_{\mathrm{grav}}= \frac{\Delta x q F}{M^{\mathrm{fu}}},
     \label{eq:capacity}
\end{equation}

where $\Delta x$ is the number of inserted cations per formula unit, $q$ is the charge of these cations in units of the elementary charge $e$ (e.g., $q = 1$ for Li$^+$ and $q = 2$ for Ca$^{2+}$), $F$ is the Faraday constant, and $M^{\mathrm{fu}}$ is the molar mass per formula unit of the intercalated material ($S^{n=0}_\mathrm{ML}$). In Figure \ref{fig:disc:screen}, the predicted volume changes and theoretical capacities are displayed for the screened structure pairs. While the volume change distributions show no distinct difference between monovalent and divalent cations, the capacity distributions reveal that divalent cations exhibit a higher density than monovalent cations at higher capacities.

\subsubsection{Evaluation of LVC intercalation structures}\label{sec:screen:disc}

For application purposes, LVC materials with high capacities are especially interesting. Therefore, we selected structure pairs with capacities above 150~mAh/g in the $|\Delta V_\mathrm{ML}| < 2.65\%$ range for further investigation and validation with DFT calculations. This selection criterion resulted in 5824 structure pairs \{$S^{\mathrm{host}}_\mathrm{DFT}$,$S^{\mathrm{final}}_\mathrm{ML}$\}. To validate the model and the predicted low volume changes of those structure pairs, we calculated their volumes with DFT. We applied the same input parameters that were used to calculate the MP data (all details are given in section S1 of the SI), except for the Brillouin zone integration method. Gaussian smearing was employed because the additionally intercalated cations sometimes introduced new symmetries, and with the default \textit{k}-point meshes determined by Pymatgen, the tetrahedron method could in some cases not be applied, because at least four symmetry-irreducible \textit{k}-points were needed for this method. For consistency in volume relaxations, we recalculated the corresponding host structures using the same computational settings. When comparing values for energies above the convex hull, we found a mean difference of 2~meV/atom relative to the MP data \cite{ongMaterialsApplicationProgramming2015,jainFormationEnthalpiesMixing2011}. The deviations indicate that MP energies provide a reasonable basis for calculating the energies above the convex hull of the newly computed intercalated structures.

Within the 5824 structure pairs that satisfy $|\Delta V_\mathrm{ML}| < 2.65\%$ in our screening workflow, 287 actually exhibited LVC behavior with $|\Delta V_\mathrm{DFT}| < 1\%$, yielding a precision of 0.05. For comparison, we also computed the volume changes of the same 5824 pairs using the sum of ionic radii as bond-length estimate. This approach identified 512 structure pairs within $|\Delta V_\mathrm{Ionic}| < 2.65\%$, of which only 12 actually satisfied the LVC criterion ($|\Delta V_\mathrm{DFT}| < 1\%$) after verification with the DFT calculation. Thus, our proposed workflow using $\mathcal{M}_{\mathrm{Bond}}$ to calculate bond lengths identifies a substantially larger number of LVC candidates in the same subset and with higher precision. To further assess the workflow's performance, we randomly sampled an additional 1000 structure pairs outside the selected range (also with capacities above 150~mAh/g) to estimate how many LVC structures are presumably  excluded by the $|\Delta V_\mathrm{ML}| < 2.65\%$ threshold. Among these 1000 pairs, we found 6 that exhibited $|\Delta V_\mathrm{DFT}| < 1\%$. Assuming this random subset is representative of the full dataset outside the threshold range, this extrapolates to approximately 1457 LVC structure pairs among the 242,887 pairs excluded. Combined with the 287 correctly identified structure pairs within the threshold, the total estimated number of LVC pairs is 1744 out of 248,711 structures. Random selection from the entire dataset would therefore yield a precision of below 0.01, demonstrating that the proposed workflow is approximately 8 times more efficient than an approach without prior filtering. The estimated recall of our workflow is 0.16 (287 out of 1744 LVC structure pairs), suggesting that further optimization of the workflow may improve the coverage.

To estimate if there was systematic under- or overestimation, we analyzed the sign disagreements between predicted and calculated volume changes: 523 (8.9\%) structures show such disagreements, with sign disagreement rates of 43.2\%, 14.1\%, and 8.4\% within the predicted ranges of $|\Delta V_\mathrm{ML}|$ of 0-–1\%, 1–-2\%, and 2–-2.65\%, respectively. This distribution indicates that sign disagreements are not systematic but rather result from model inaccuracy near low volume changes, where small absolute errors can flip the sign, and for the primary objective of identifying LVC materials, the sign is of secondary importance. Since the Model $\mathcal{M}_{\mathrm{Vol.}}$ does not explicitly account for changes in lattice angles, we analyzed how frequently such distortions occur in the DFT validation set. The median maximum angle change (i.e., the largest of $\Delta \alpha$, $\Delta \beta$, $\Delta \gamma$ per structure pair) was $1.57^\circ$, and only 140 out of 5824 structure pairs exhibited angle changes larger than $5^\circ$, indicating that angular distortions are relatively rare. While incorporating angle variations could further refine the model, the largest deviations are more plausibly attributed to inaccuracies in the predicted bond lengths, for example when certain local environments or specific oxidation-state combinations are underrepresented in the training data. Although the screening is not fully efficient, it nevertheless represents a valuable first-step filter capable of rapidly reducing large candidate sets.

Table \ref{tab:newLVCmaterials} presents selected structure pairs for which the DFT validation calculations yielded a low volume change. The table lists theoretical metrics for battery applications, namely the theoretical capacity [Eq. (\ref{eq:capacity})] and the theoretical intercalation voltage, which is calculated as follows:

\begin{equation}
	U = -\frac{E_1-E_0- \Delta x E_M}{\Delta x e}.
	\label{eq:Spannung}
\end{equation}
$E_{0}$ and $E_{1}$ are the energies of the unit cells with low and high cation concentrations, respectively, $E_{M}$ is the energy per atom of the metallic phase of the intercalated cation species, $\Delta x$ is the number of intercalated cations per cell, and $e$ is the elementary charge. Using metallic references, we calculated the voltage against M/M$^{+}$ or M/M$^{2+}$. We used reference energies from MP for the body-centered cubic structures for Li, Na, and K, the hexagonal close-packed structure for Mg, and the face-centered cubic structure for Ca. For the calculation of the energy above the convex hull for the intercalated structures, we also used energies from the Materials Project. In the following, we discuss our workflow using a selection of materials, highlighting areas where improvements are likely and indicating directions for further research.

The insertion of Li, Na and Mg into $\alpha$-V$_2$O$_5$ has been experimentally investigated, and it was found that phase transformations occur at room temperature, which limit the achievable intercalation levels \cite{whittinghamRoleTernaryPhases1976,niuStructureEvolutionV2O52023a,mukherjeeDirectInvestigationMg2017}. In this case, the solid-solution assumption is not valid, and a different kind of simulation would be required to capture such phase transformations. The energy above the convex hull value for this composition indicates that another phase or phase decomposition route exist, which is thermodynamically favorable. The insertion of Ca into $\alpha$-V$_2$O$_5$ has also been studied in detail by Wang \textit{et al.}, who observed that the process occurs in three distinct stages \cite{wangRobustVanadiumPentoxide2016}. Although multi-stage insertion was not considered in the present work, where the corresponding Wyckoff positions were filled completely in one step, this is readily feasible within the framework of our workflow. Stepwise cation insertion would allow the calculation of intermediate states and enable the screening for materials that maintain low volume changes throughout the entire intercalation range or through parts of it. The structures considered here often have multiple vacant Wyckoff positions, and in this framework, we have investigated the occupation of one site at a time. More generally, one could use a host structure, insert a set of cations, and then treat the resulting intercalated structure as a new host for further cation insertion. The volume change can be calculated at each step, providing a detailed map of structural evolution throughout the intercalation process.

For Mg insertion into VF$_5$, several distinct intercalation sites had a LVC behavior at different voltages. The grid-based search strategy employed here enables a systematic exploration of the intercalation configuration space. However, it does not guarantee identification of the energetically most favorable intercalation sites. This approach could be further improved to identify the most promising intercalation sites more efficiently, as demonstrated, for example, by the charge-density-based methodology proposed by Shen et al. \cite{shenChargedensitybasedGeneralCation2020}. However, VF$_5$ as a host structure presents also fundamental practical limitations, because it is only solid below 20~$^\circ$C  and turns gaseous above 50~$^\circ$C, rendering it unsuitable for battery applications \cite{Wiberg+2008}. Such temperature stability issues are beyond the capability of our workflow and are not inherent to DFT data taken from the MP database. One property that can be assessed is the energy above the convex hull, which compares the energy of a given phase to the most stable combination of competing phases. For Mg$_2$K$_6$Co$_2$O$_7$ and Li$_2$Rb$_2$CrF$_6$, these values are relatively high, indicating that these materials are thermodynamically unstable with respect to decomposition, suggesting that these compounds are unlikely to be suitable for practical applications.

\begin{landscape}
\begin{table}
\centering
\caption{Selected structure pairs with low volume changes according to the presented workflow and DFT calculations. The Wyckoff positions are given for the intercalated cations in the DFT relaxed structures. In addition, theoretical metrics important for practical battery applications are provided. The space group is the same for the host and occupied structures.}
\centering
\begin{tabular}{llll
    S[table-format=-1.3]
    S[table-format=3.1]
    S[table-format=1.2]
    S[table-format=1.3]}
\hline
$S^{\mathrm{host}}_\mathrm{DFT}$ & $S^{\mathrm{occ.}}_\mathrm{DFT}$ & Space group & Wyckoff position & 
{$\Delta V_\mathrm{DFT}$ [\%]} & {Capacity [mAh/g]} & {Voltage [V]} & {$E_{\mathrm{Hull}}^{\mathrm{occ.}}$ [eV/atom]} \\
\hline
ZrV$_2$O$_7$ & Ca$_2$ZrV$_2$O$_7$ & $Pa\bar{3}$ & 8c (0.801, 0.801, 0.801) & -0.238 & 278.3 & 0.67 & 0.015 \\
V$_2$O$_5$ & Li$_2$V$_2$O$_5$ & $Pnma$ & 4a \& 4b & 0.381 & 273.8 & 2.56 & 0.163 \\
V$_2$O$_5$ & MgV$_2$O$_5$ & $Pnma$ & 4c (0.426, 0.250, 0.326) & -0.595 & 260.0 & 1.06 & 0.102 \\
LiVF$_5$ & MgLi$_2$V$_2$F$_{10}$ & $Pnma$ & 4c (0.788, 0.250, 0.227) & -0.535 & 243.6 & 1.90 & 0.072 \\
ZnCr$_2$F$_{12}$ & Mg$_2$ZnCr$_2$F$_{12}$ & $P\bar{1}$ & 2i (0.790, 0.721, 0.504) & 0.429 & 240.4 & 2.70 & 0.029 \\
V$_2$O$_5$ & Na$_2$V$_2$O$_5$ & $Pmmn$ & 4e (0.250, 0.894, 0.489) & 0.005 & 235.2 & 2.25 & 0.121 \\
K$_6$Co$_2$O$_7$ & Mg$_2$K$_6$Co$_2$O$_7$ & $P2_1/c$ & 4e (0.412, 0.566, 0.867) & -0.013 & 209.0 & 0.74 & 0.115 \\
CsCuO$_2$ & CaCsCuO$_2$ & $Cmcm$ & 4c (0.000, 0.564, 0.250) & -0.959 & 199.6 & 0.23 & 0.000 \\
VF$_5$ & LiVF$_5$ & $P2_1/c$ & 4a (0.610, 0.345, 0.195) & 0.367 & 175.3 & 4.68 & 0.035 \\
Sc$_2$(MoO$_4$)$_3$ & Mg$_2$Sc$_2$(MoO$_4$)$_3$ & $Pbcn$ & 8d (0.241, 0.924, 0.309) & -0.915 & 173.4 & 0.31 & 0.000 \\
VF$_5$ & MgV$_2$F$_{10}$ & $P2_1/c$ & 4e (0.211, 0.248, 0.965) & 0.577 & 169.5 & 3.21 & 0.000 \\
VF$_5$ & MgV$_2$F$_{10}$ & $P2_1/c$ & 4e (0.296, 0.248, 0.036) & -0.969 & 169.5 & 3.19 & 0.000 \\
Mn(ReO$_4$)$_2$ & Ca$_2$Mn(ReO$_4$)$_2$ & $P\bar{3}$ & 2d (0.333, 0.667, 0.862) & -0.490 & 168.7 & 7.11 & 0.000 \\
FePO$_4$ & CaFe$_2$(PO$_4$)$_2$ & $P\bar{3}$ & 1a (0.000, 0.000, 0.325) & -0.906 & 168.7 & 1.67 & 0.021 \\
CuPO$_4$ & LiCuPO$_4$ & $Pnma$ & 4c (0.100, 0.250, 0.036) & 0.126 & 162.0 & 3.40 & 0.000 \\
VF$_5$ & CaV$_2$F$_{10}$ & $P2_1$ & 2a (0.488, 0.676, 0.366) & 0.111 & 161.5 & 3.31 & 0.046 \\
KV$_3$O$_8$ & Li$_2$KV$_3$O$_8$ & $P2_1/m$ & 4f (0.674, 0.090, 0.077) & 0.580 & 160.6 & 2.31 & 0.000 \\
VPO$_5$ & MgV$_2$(PO$_5$)$_2$ & $P2_1/c$ & 4e (0.548, 0.206, 0.428) & -0.136 & 154.0 & 1.91 & 0.044 \\
Rb$_2$CrF$_6$ & Li$_2$Rb$_2$CrF$_6$ & $Fmmm$ & 8e & 0.109 & 152.8 & 1.61 & 0.199 \\
V$_2$Cu$_2$O$_7$ & Li$_2$V$_2$Cu$_2$O$_7$ & $C2/c$ & 8f (0.056, 0.998, 0.392) & -0.767 & 151.1 & 2.69 & 0.075 \\
\hline
\end{tabular}
\label{tab:newLVCmaterials}
\end{table}
\end{landscape}

KV$_3$O$_8$ has previously been investigated as a possible Li insertion material and was found to exhibit capacities comparable, but slightly lower than the theoretical values calculated here \cite{fengSimpleMethodSynthesis2013}. The material was studied in nanostructured form, and no results regarding volume changes upon lithiation were reported. For $\alpha$-V$_2$Cu$_2$O$_7$, lithiation was investigated experimentally \cite{eguchiLithiumCuVO1993}. Over the range $x < 1$, no changes in lattice spacing were observed, consistent with the LVC behavior calculated for the broader composition range of $0 < x < 2$. In contrast, Ca$_x$ZrV$_2$O$_7$, Mg$_x$Li$_2$V$_2$F$_{10}$, Mg$_x$ZnCr$_2$F$_{12}$, Ca$_x$CsCuO$_2$,  Mg$_2$Sc$_2$(MoO$_4$)$_3$, Ca$_x$Mn(ReO$_4$)$_2$, Ca$_x$Fe$_2$(PO$_4$)$_2$, Li$_x$CuPO$_4$, Mg$_x$V$_2$(PO$_5$)$_2$, and Li$_x$KV$_3$O$_8$ in the crystal structures investigated here, have not yet been studied as intercalation materials to the best of our knowledge. Most of these compounds exhibit relatively low intercalation voltages, which is advantageous for anode applications, where low potentials are desirable. For cathode applications, however, higher operating voltages would be necessary. If such higher voltages could be achieved, for example through alternative intercalation sites as discussed above, some of these materials, particularly ZrV$_2$O$_7$ with its high capacity, could become especially attractive as cathodes. Overall, the combination of LVC behavior, moderate to high capacities and voltages suggests that these materials are promising candidates for experimental investigation to assess their practical electrochemical performance.

The workflow successfully enables filtering of large datasets and has delivered several new structure pairs with low volume changes during (de-)intercalation. Future extensions may include application to crystal structures beyond \textit{3d}-TM oxides and fluorides, provided that sufficiently consistent datasets are available for training and validating the model $\mathcal{M}_{\mathrm{Bond}}$. Improvements in bond-length predictions and refined site-selection strategies may further strengthen this approach as a powerful step in the discovery of new active electrode materials. Potential improvements may e.g., include the incorporation of descriptors from next-nearest neighbors to better capture longer-range structural effects or the inclusion of magnetic properties, such as the magnetic moments of atoms in the descriptor set.

\section{Conclusion}\label{sec:mod:conclusion}

In this work, we present a workflow for predicting the volume change of potential electrode materials upon intercalation of ions. We demonstrate and validate this approach for \textit{3d}-TM oxides and fluorides using a dataset of DFT results from the MP database. The workflow starts by placing ions into a host structure and uses a gradient boosting model to calculate the bond lengths between neighboring atoms in this intercalated structure. These calculated bond lengths are used to iteratively adjust the structure until a self-consistency criterion is met. Finally, the volume of the final structure allows to estimate the volume change of the considered structure pair under ion intercalation. The estimated bond lengths are in good agreement with DFT data (MAE = 0.02~\AA). Using these bond lengths results in a median absolute error of 2.04\% in the predicted volume changes compared to volume changes calculated with DFT. Improving bond length accuracy would further reduce errors in volume predictions and, consequently, enhance the efficiency of the workflow. We have applied the workflow to a large-scale screening of \textit{3d}-TM oxides and fluorides. Using DFT calculations, we validated the most promising candidates selected by our screening approach and identified 287 LVC structure pairs. As a filtering tool of large datasets, the workflow is approximately 8 times more efficient than random sampling and successfully identifies 24 times more LVC structures than when using tabulated ionic radii for the bond-length estimation. Within the chemical space defined by the training data, the workflow follows an effective strategy for pre-screening large candidate sets. It substantially reduces the number of DFT calculations needed to discover promising low-volume-change intercalation materials.

\section*{Data availability}
The code supporting the findings of this study, including scripts for structure generation, feature extraction, and volume prediction, and the trained XGBoost model are openly available at \cite{VolumePredictionWorkflow}. The DFT data used in this study were obtained from the Materials Project \cite{jainCommentaryMaterialsProject2013} and are therefore not included with this repository, as they are publicly accessible through the original source.

\section*{Supporting Information}
Computational details of the workflow and DFT calculations; detailed description of the volume prediction model; illustrative application to the system FePO$_4 \leftrightarrow$ LiFePO$_4$


\section*{Acknowledgments}
This work was funded by the German Research Foundation (DFG, Grant No.\ EL 155/29-1). The authors acknowledge support by the state of Baden-Württemberg through bwHPC and the German Research Foundation (DFG) through grant no INST 40/575-1 FUGG (JUSTUS 2 cluster). Crystallographic drawings were created with the software VESTA \cite{mommaVESTA3Threedimensional2011}.\\

\newpage

\setcounter{page}{1}
\setcounter{section}{0}
\setcounter{figure}{0}
\setcounter{table}{0}
\setcounter{equation}{0}

\renewcommand{\thepage}{S\arabic{page}}

\renewcommand{\thesection}{S\arabic{section}}
\renewcommand{\thesubsection}{S\arabic{section}.\arabic{subsection}}
\renewcommand{\thesubsubsection}{S\arabic{section}.\arabic{subsection}.\arabic{subsubsection}}

\renewcommand{\thefigure}{S\arabic{figure}}

\renewcommand{\thetable}{S\arabic{table}}

\renewcommand{\theequation}{S\arabic{equation}}

\begin{center}
    \vspace*{2cm}
    {\LARGE\bfseries Supporting Information: High-Throughput-Screening Workflow for Predicting Volume Changes by Ion Intercalation in Battery Materials}
    
    \vspace{1cm}
    
\end{center}

\vspace{1cm}
\section{Computational details}\label{sec:app:comp_details}

To determine the neighbors of an atom in a crystal structure, the \textit{CrystalNN} tool of the Python package \textit{pymatgen} was used, which was developed specifically for periodic structures \cite{ongPythonMaterialsGenomics2013}. The local structure order parameters were calculated for each atom using the Python package \textit{matminer} \cite{wardMatminerOpenSource2018}. Oxidation states were assigned using the procedure implemented in the Materials Project \cite{jainCommentaryMaterialsProject2013}. The different possible oxidation states are compared to entries in the \textit{Inorganic Crystal Structure Database} (ICSD) to select the most probable assignments \cite{zagoracRecentDevelopmentsInorganic2019}. Oxidation states are used as one of only two elemental descriptors in the ML model and can therefore impact the bond-length prediction strongly. However, restricting the dataset to oxides and fluorides, where large datasets exist, provides confidence in the assigned states.

Hyperparameters of the gradient boosting model were optimized using the Python package \textit{Optuna} \cite{akibaOptunaNextgenerationHyperparameter2019a}. The final model, for which the results are presented, and which was used in the screening procedure, used a maximum tree depth of 10, a learning rate of 0.05, 600 trees, a subsample ratio for trees of 0.4, and a L2 regularization weight of 19. For all other parameters the default values of XGBoost were used.

All density functional theory (DFT) calculations were performed using the Vienna \textit{Ab initio} Simulation Package (VASP) \cite{kresseEfficientIterativeSchemes1996} with the projector-augmented wave (PAW) method \cite{kresseUltrasoftPseudopotentialsProjector1999}. The Perdew-Burke-Ernzerhof (PBE) generalized gradient approximation (GGA) was employed for the exchange-correlation functional \cite{perdewGeneralizedGradientApproximation1996}. A plane-wave cutoff energy of 520~eV and a \textit{k}-point grid corresponding to a reciprocal-space density of $\sim$64~\AA$^{-3}$ were used. Structural relaxations were carried out with an energy convergence criterion of $5 \times 10^{-5}$~eV per atom using the conjugate gradient algorithm, allowing both atomic positions and cell shape/volume to relax. The calculations were performed with spin polarization. Gaussian smearing with a width of 0.05~eV was used for Brillouin zone intergration. Strongly correlated electron effects were treated using the DFT$+U$ approach in the effective $U_{\mathrm{eff}}$ = ($U-J$) scheme \cite{dudarevElectronenergylossSpectraStructural1998}, with the values of $U_{\mathrm{eff}}$ for these TM elements: 3.25~eV for V, 3.7~eV for Cr, 3.9~eV for Mn, 5.3~eV for Fe, 3.32~eV for Co, 6.2~eV for Ni and 6.2~eV for W.

\section{Detailed computational procedure for the volume prediction ($\mathcal{M}_{\mathrm{Vol.}}$)}\label{sec:app:MVOL_detail}

In this section, we present the detailed formalism of the model $\mathcal{M}_{\mathrm{Vol.}}$, which iteratively adjusts a structure (i.e., the atomic positions and lattice parameters) to match the predicted equilibrium bond lengths. In each iteration $n$, the difference between the bond lengths in the current structure $S^{n}_\mathrm{ML}$ and the model-predicted bond lengths $l^{\mathrm{pred.}}_{ij}$ is calculated:

\begin{equation}
    \delta l^n_{ij} = l^n_{ij} - l^{\mathrm{pred.}}_{ij}.
\label{eqn:bond_err}
\end{equation}

In addition, the direction of the bond between atoms $i$ and $j$ in the structure $S^{n}_\mathrm{ML}$ along the lattice parameters is determined by the difference in the fractional coordinates of the two atoms:

\begin{equation}
   \begin{bmatrix}dx_{ij}\\dy_{ij}\\dz_{ij} \end{bmatrix}^n = \begin{bmatrix}x_i\\y_i\\z_i  \end{bmatrix}^n - \begin{bmatrix}x_j\\y_j\\z_j  \end{bmatrix}^n.
\label{eqn:bond_dir}
\end{equation}

Using those differences to all neighboring atoms $j$, the coordinates of atom $i$ are changed in the direction of the predicted values. This is done by the following rule:

\begin{equation}
    \begin{bmatrix}x_i\\y_i\\z_i  \end{bmatrix}^{n+1} = \begin{bmatrix}x_i\\y_i\\z_i  \end{bmatrix}^{n} + \beta \sum_{j} \delta l^n_{ij} \begin{bmatrix}dx_{ij}\\dy_{ij}\\dz_{ij} \end{bmatrix}^n
\label{eqn:atom_update}
\end{equation}

\noindent with the sum running over all neighbors $j$, and $\beta$ is a parameter to regulate the strength of the change.

Since, obviously, the change in volume of the unit cell is caused by a change in the lattice parameters, the following rule is used to calculate the lattice parameters for the next step:

\begin{align}
\begin{bmatrix}a_1&a_2&a_3\\b_1&b_2&b_3\\c_1&c_2&c_3 \end{bmatrix}^{n+1} 
&= \begin{bmatrix}a_1&a_2&a_3\\b_1&b_2&b_3\\c_1&c_2&c_3 \end{bmatrix}^{n} \nonumber \\
&\quad \cdot \left(1 - \gamma \sum_{ij\text{-pairs}} \delta l^n_{ij}
\begin{bmatrix}|dx_{ij}|\\|dy_{ij}|\\|dz_{ij}| \end{bmatrix}^n\right)
\label{eqn:lattice_update}
\end{align}

\noindent where the parameter $\gamma$ regulates the magnitude of the adjustment. In this formulation, directions for which the sum of the differences $\delta l^n_{ij}$ is positive (i.e., the current bond lengths exceed the predicted values) are scaled down, while directions for which $\delta l^n_{ij}$ is negative (current bond lengths are shorter than predicted) are scaled up. This operation modifies only the lengths of the lattice vectors, without changing the cell shape; the original lattice angles are preserved in the updated cell. The volume of the updated cell is then calculated, and if the change relative to the previous iteration $n$ falls below a predefined threshold, the structure $S^{n+1}_\mathrm{ML}$ is accepted as final and the volume change is computed relative to the initial structure $S^{n=0}_\mathrm{ML}$. Otherwise, the procedure is repeated starting from Eq.~(\ref{eqn:bond_err}), with the bond lengths $l^n_{ij}$ now updated.

With the bond lengths being calculated in Angstrom, the parameters for the volume-change calculations were set as $\beta = 0.001$/\AA~and $\gamma = 0.003$/\AA. The self-consistency criterion was defined as the convergence of the total cell volume and was set to $10^{-3}$~\AA$^3$, meaning the iteration loop stops when the volume difference between consecutive steps falls below this value. We set the maximum number of iterations to 500. Across the training dataset of 2707 structure pairs, 72.6\% converged successfully, with most converged structures reaching convergence in fewer than 150 iterations (mean = 147 iterations), while 27.4\% reached the maximum iteration limit. No structures failed due to errors. Increasing the iteration limit to 1000 improved the convergence rate to 94.2\%, with only 5.8\% reaching the limit. However, the accuracy remained nearly unchanged (median absolute error (n=1000): 2.056\% vs.\ median absolute error (n=500): 2.042\%), demonstrating that even with more iterations, the final volumes do not differ significantly. This supports the approach of using the volume from the last iteration when convergence is not reached, and justifies the choice of 500 iterations as a reasonable compromise between computational efficiency and accuracy for the high-throughput screening.

\section{Application of $\mathcal{M}_{\mathrm{Vol.}}$ to FePO$_4 \leftrightarrow$ LiFePO$_4$}\label{sec:app:MVOL_example}

\begin{figure}[H]
\centering
\includegraphics[width=0.5\columnwidth]{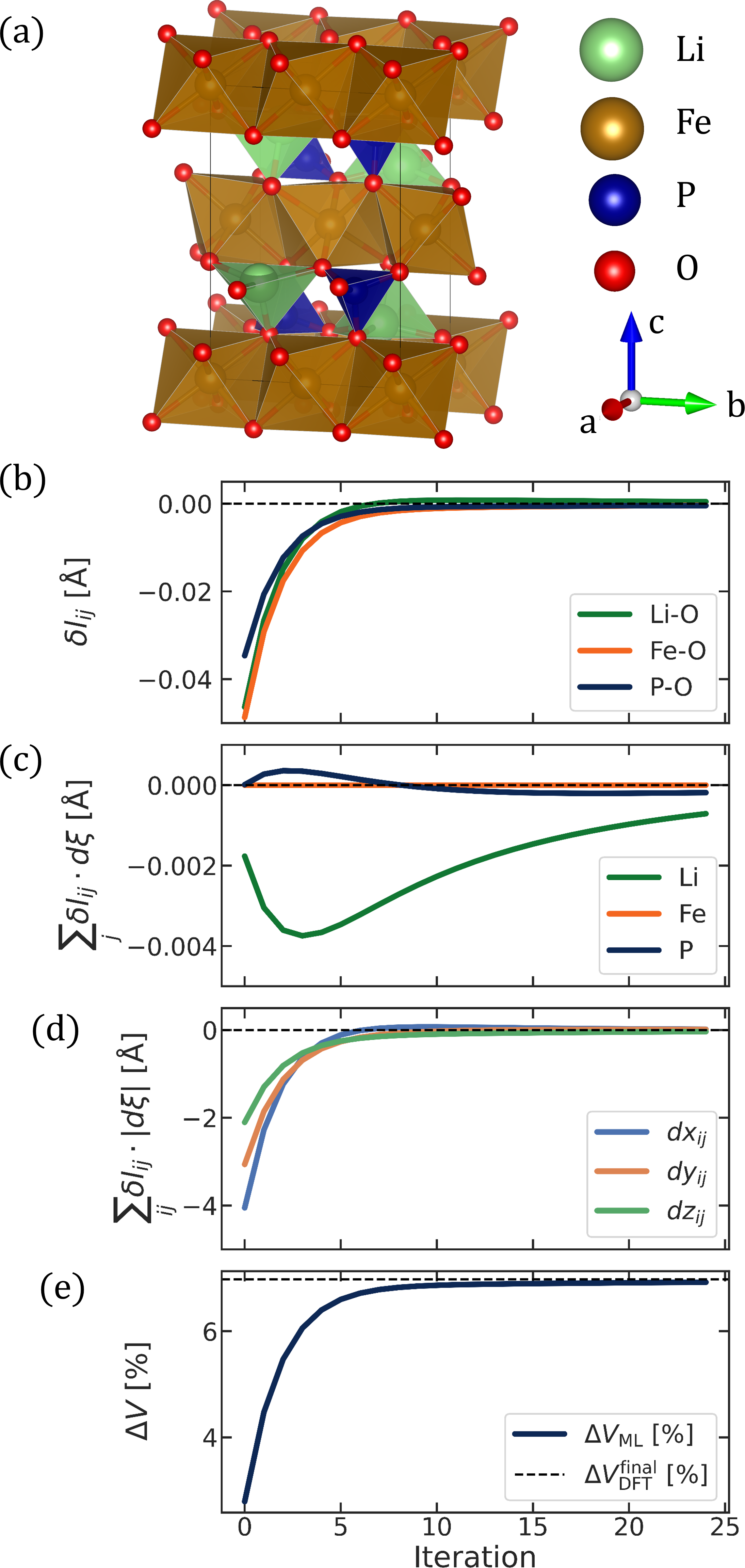	}
\caption{(a): Unit cell of LiFePO$_4$ with space group \textit{Pnma}. (b)-(e): Selected calculated quantities during the iterative adjustment of the LFPO structure over 25 iterations: (b) differences between current and predicted bond lengths (three selected bonds are shown); (c) sum of differences of all neighboring atoms $j$ for three selected atoms, multiplied by the bond direction [Eq. (\ref{eqn:bond_dir})], (d) sum over all differences of neighboring ions multiplied by the absolute bond direction values for all bonds [see Eq. (\ref{eqn:lattice_update})], (e) volume change of the unit cell, with the DFT value indicated as a horizontal line.}
\label{fig:LFPO_example}
\end{figure}

We illustrate the iterative procedure of the model $\mathcal{M}_{\mathrm{Vol.}}$ for the example of the structure pair FePO$_4 \leftrightarrow$ LiFePO$_4$ (LFPO), using the predicted bond lengths from the model $\mathcal{M}_{\mathrm{Bond}}$ as input for $l^{\mathrm{pred.}}_{ij}$ [see Eq.\ (\ref{eqn:bond_err})]. The unit cell of LFPO with space group \textit{Pnma} is shown in Fig.~\ref{fig:LFPO_example} (a). The difference between $l^{\mathrm{pred.}}_{ij}$ and $l^{n}_{ij}$ of three exemplarily chosen bonds are plotted in Fig.~\ref{fig:LFPO_example} (b). For these bonds, the predicted lengths $l^{\mathrm{pred.}}_{ij}$ are larger than the actual bonds $l^{n=0}_{ij}$ in $S^{n=0}_\mathrm{ML}$, resulting in negative bond errors [Eq.\ (\ref{eqn:bond_err})]. The intial difference is more pronounced for bonds involving the electrochemically active Fe ion (Fe-O) and the newly inserted Li ion (Li-O), compared to bonds of P with O. Using these differences and the bond directions, the atomic coordinates are updated iteratively. For the Fe ion, which is nearly perfectly octahedrally coordinated by O ions, vectors in opposite directions cancel each other [see Eq.\ (\ref{eqn:bond_dir})], resulting in the sum over all neighbors being close to zero [see Fig.~\ref{fig:LFPO_example} (c)]. In contrast, for Li and P ions, the sum is nonzero which leads to an adjustment of their coordinates.

To adjust the lattice parameters, the sum of all differences $\delta l^{n}_{ij}$ multiplied by the absolute values of the bond directions for all bonds is used. If this sum is negative for a given direction, the current bonds are shorter than predicted, and the lattice parameter along that direction is stretched according to Eq.~(\ref{eqn:lattice_update}). The three sums along the three lattice directions are shown in Fig.~\ref{fig:LFPO_example} (d). In LFPO, all lattice parameters are increased during the iterative procedure, which results in a progressive increase of the unit cell volume [Fig.~\ref{fig:LFPO_example} (e)]. For comparison, the DFT value of the volume change, 6.98\%, is indicated. The calculated volume change of 6.96\% by the workflow closely matches the DFT result.

\bibliography{references}

\end{document}